\documentclass[12pt]{article}
\usepackage{psfig,epsfig}
\def\bd{\begin{document}} \def\ed{\end{document}}
\def\emp{\end{minipage}} \def\bmp{\begin{minipage}}
\def\bcc{\begin{center}} \def\ecc{\end{center}} \def\npg{\newpage}
\def\beq{\begin{equation}} \def\eeq{\end{equation}} \def\hph{\hphantom}
\def\hs{\hskip} \def\vs{\vskip} \def\hf{\hfill} \def\ej{\vfill\eject}
\def\cl{\centerline} \def\ob{\obeylines}  \def\ls{\leftskip}

\begin{document}

\vs 4cm
\centerline{\LARGE On the Stability of Rapidity Gap Analysis}

\vs 0.5cm
\centerline{\large Liao Hongbo \ \ \ Wu Yuanfang}

\centerline{Institute of Particle Physics, Huazhong Normal University, Wuhan 430079 China}

\vs 3cm
\centerline{\bf Abstract}

\qquad
It is argued that the newly introduced moments of rapidity gaps for the 
event-by-event fluctuations depends on the number of events and multiplicity. 
The interesting ones of them are unstable under ISR energies of h-h collisions. 
The instability get well improved when multiplicity increases.

\newpage

\qquad
In recent decade, the event-by-event fluctuations become more and more 
important in multiparticle production and relativistic heavy ion collisions. 
This is partly due to the fact that the large local multiplicity fluctuations have been
observed in all kind of collisions~\cite{wolfram} and that 
the appearance of new state of matter, Quark Gluon Plasma (QGP), predicted by QCD, will 
definitely associate with large energy density fluctuations~\cite{qgp}. How to measure 
the fluctuations turn to be a powerful tool in probing the QGP phase transition.
However, in the market at present, there is no good suggested measure for the
purpose. In order to measure first the event-by-event fluctuations of low multiplicity
samples, such as those in h-h collisions under ISR energies, R.C.Hwa and Q.Zhang newly 
introduced the rapidity gap analysis~\cite{hwa1} after a long exploration for the 
end~\cite{hwa2}. Though it has been used to analyze experimental data ~\cite{wang}, the
efficiency of the method for low energies of h-h collisions has not been seriously checked. 
In this letter, we are going to study the efficiency of the method.

\qquad The rapidity gap is defined by the difference of rapidity of two neighbor particles 
in an event,

\begin{equation}
x_{i}=X_{i+1}-X_{i},
\hs 1cm i=0,\cdots,N,
\end{equation}

\noindent where $X_i$ is the cumulant variable of rapidity of $i$th particle, which is free of the 
influence of energy conservation in rapidity distribution~\cite{cumu} and is uniformly 
distributed from $0$ to $1$. $N$ is the total number of particles in the event. 
The set $S_{e}$ of $N+1$ number: $S_{e}=\{x_{i}\mid i=0,\cdots,N\}$ provides information of 
rapidity distribution of all particles in the event. A quantitative character of single event 
therefore can be estimated by the moments of its rapidity gaps:

\begin{equation}
G_{q}=\frac{1}{N+1}\sum_{i=0}^{N}x_{i}^q,
\end{equation}

\noindent or: 

\begin{equation}
H_{q}=\frac{1}{N+1}\sum_{i=0}^{N}(1-x_{i})^{-q},
\end{equation}

\noindent where moment order $q$ is an integer. $G_{q}$ and $H_{q}$ vary from event to event.
R. C. Hwa and Q. Zhang suggest the simplest moments of them: 

\begin{equation}
s_{q}=-\langle G_{q}\ln G_{q}\rangle,
\end{equation}

\noindent and

\begin{equation}
\sigma_{q}=\langle H_{q}\ln H_{q}\rangle
\end{equation}

\noindent as the descriptions of event-by-event fluctuations of the sample.
Here, $\rangle\dots\langle$ is the average over all the events in the sample. 
However, according to this description, the statistical fluctuations, 
which we are not interested in, are included. 
In order to reduce the statistic fluctuations, they further suggested the same estimation 
for pure statistical sample as

\begin{equation}
s_{q}^{st}=-\langle G_{q}^{st}\ln G_{q}^{st}\rangle,
\end{equation}

\noindent and

\begin{equation}
\sigma_{q}^{st}=\langle H_{q}^{st}\ln H_{q}^{st}\rangle
\end{equation}

\noindent and defined 

\begin{equation}
S_{q}=\frac{s_{q}}{s_{q}^{st}},
\end{equation}

\noindent and

\begin{equation}
\Sigma_{q}=\frac{\sigma_{q}}{\sigma_{q}^{st}}
\end{equation}

\noindent as interesting measure of event-by-event fluctuations of the sample,
{\it i.e.}, so called the measure of erraticity in terms of rapidity gaps. 

\qquad From the definition of rapidity gaps above, it is clear that the rapidity gaps will be 
large (small) and vary violently (smoothly) from event to event if the multiplicity of 
the event is very low (high), such as h-h collisions 
under (above) ISR energies. Since rapidity gap $x_i$ is less than $1$, event moments 
$G_{q} \ll 1$ and $H_{q} \gg 1$. The measure of $\Sigma_{q}$ is simply the amplification of 
$S_{q}$. The higher the moment order is, the bigger are the time of the amplification. 
In this case, a stable measure of $\Sigma_{q}$ requires much larger
number of events than that of $S_{q}$ so that the fine structure of $H_{q}$ in the sample can
be completely demonstrated. Moreover, for $H_q$, only those events with multiplicity 
$N \ge q+1$ are available for the average. It makes the measure of $\Sigma_q$ even more unstable
in comparison to $S_{q}$ in the same sample. How the measures of $S_q$ and $\Sigma_q$
depend on the number of events and multiplicity and how to get the stable measures of 
$S_q$ and $\Sigma_q$ are the questions that we are going to answer in this letter.

\qquad The simplest way of the investigation is to simulate a statistical sample, where 
the number of events and multiplicities are all controllable. It is enough for us to estimate 
the $s_{q}^{st}$ and $\sigma_{q}^{st}$ of the sample. The cumulent variables of rapidities of
a statistical event with multiplicity $N$ are constructed by $N$ random number 
$X_{1}^{\prime}, \cdots, X_{N}^{\prime}$ which is uniformly distributed in [0, 1]. 
Here the distribution of multiplicity is taken from h-h collisions of NA22 experiments as
an example. Then according to Eqs.(6)-(7),   
$s_{q}^{st}$ and $\sigma_{q}^{st}$ can be calculated.

\qquad Firstly, the dependency of $s_{q}^{st}$ on the number of events is presented in fig.1, 
where the number of events are $10,000$, $20,000$, $\cdots$, $70,000$ and moment order is 
from $q=1$ to $8$. In these range of number of events, the behavior of all orders' 
$\ln s_{q}^{st}$ are very stable. This results tell us that the measure of $S_q$ are 
stable in experimentally allowed number of events even if the multiplicity is low.  

\qquad Then, let's turn to the same dependency of $\sigma_{q}^{st}$. The results is 
given in Fig.2(a), (b) and (c) for the orders of moment $q=1, 2, 3$ respectively.
It is clear from the results that the higher order of moment is, the larger are the number 
of events for a stable measure of $\sigma_{q}^{st}$. For $q=1$, a stable measure can be
reached at about $N_{event}=100,000$, while for $q=2$, $N_{event}$ has to go up to $500,000$,
for $q=3$, $N_{event}=1,500,000$ is out of the range of all experiments. Therefore, the 
measure of $\sigma_{q}^{st}$ is rather unstable in experimentally reachable number of 
events if multiplicity is low.  

\qquad In order to improve the measure of $\sigma_{q}^{st}$, we slightly increase 
multiplicity from above $1-25$ to $11-35$ under same multiplicity distribution as NA22 for all 
corresponding multiplicities.
The results are provided in Fig.3. Now for moment order $q$ from $1$ to $5$, the measures of 
$\sigma_{q}^{st}$ are all stable at number of events only about $10,000$. 
The stability of $\sigma_{q}^{st}$ get well improved by slightly increasing multiplicity.

\qquad From the discussions and MC results of statistical sample above, the suggested measure 
of $S_{q}$ for rapidity gaps is stable even under ISR collision energies.
However, as already pointed out in ~\cite{hwa1} that $S_{q}$ is essentially $1$, 
and therefore not very interesting. Unfortunately, another suggested measure of rapidity gaps 
$H_q$ is unstable under 
ISR collision energies. But the instability gets well improved by increasing multiplicity of
events in the sample. It means that the measure is applicable for h-h collisions above the ISR 
energies or for heavy ion collisions

\vs 3cm
{\bf Figure Captions}
\begin{itemize}

\vs 1cm
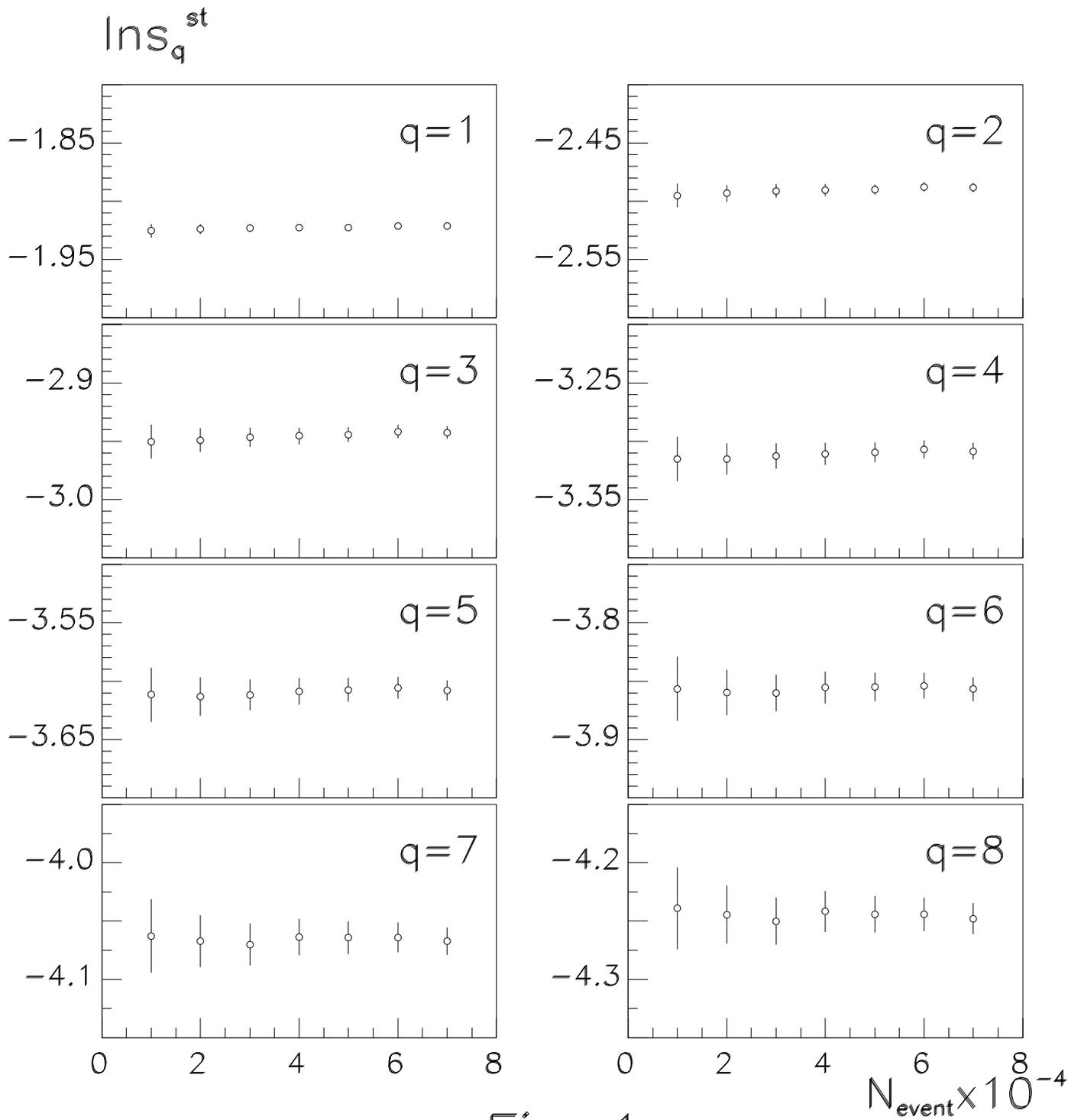
\item[\bf Fig.~1] $\ln s_{q}^{st}$ ${\sl vs.}$ $N_{event}$ for different order of moments $q$.

\vs 1cm 
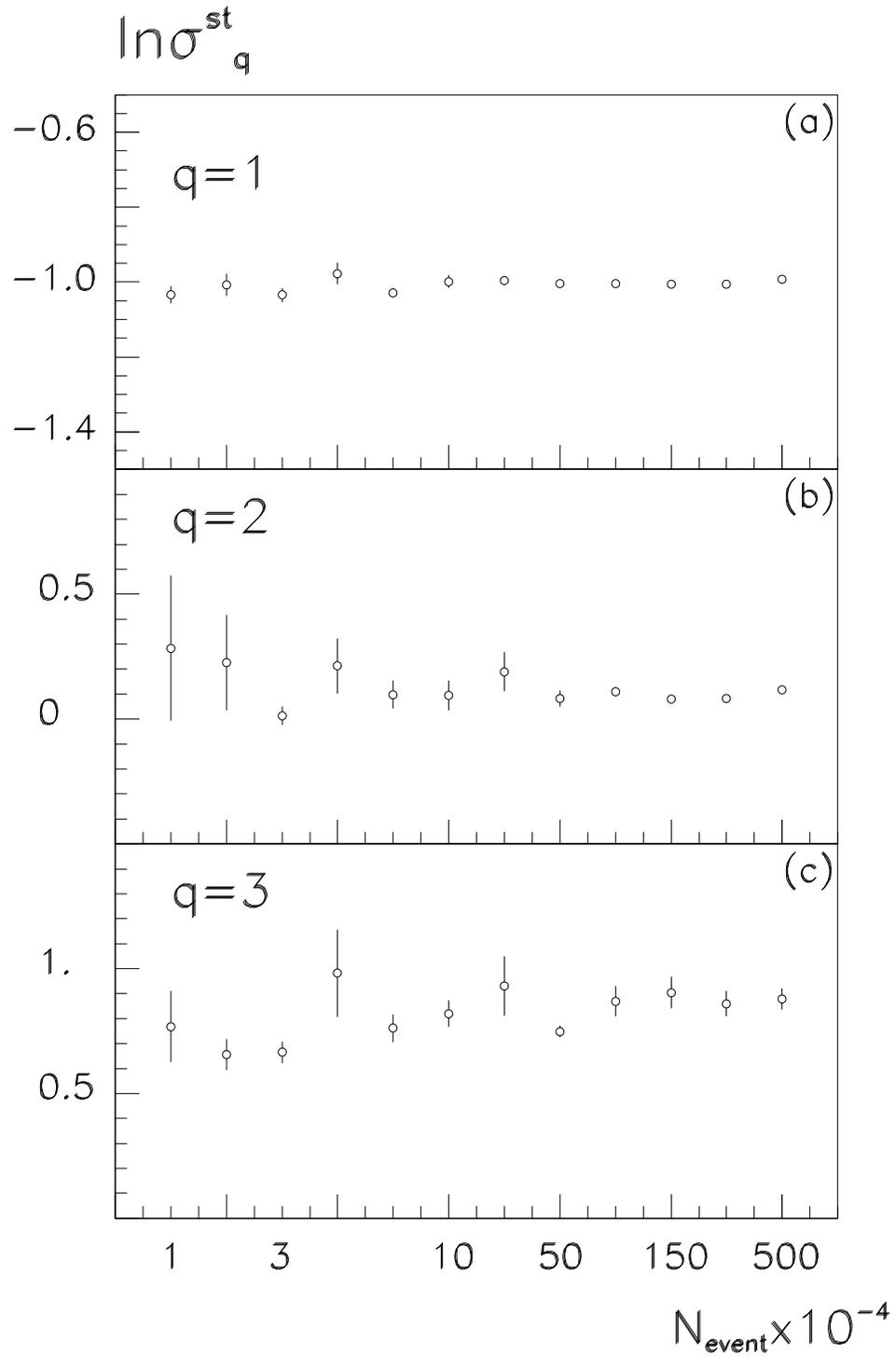
\item[\bf Fig.~2] $\ln \sigma_{q}^{st}$ ${\sl vs.}$ $N_{event}$ for different order of moments 
$q$.

\vs 1cm
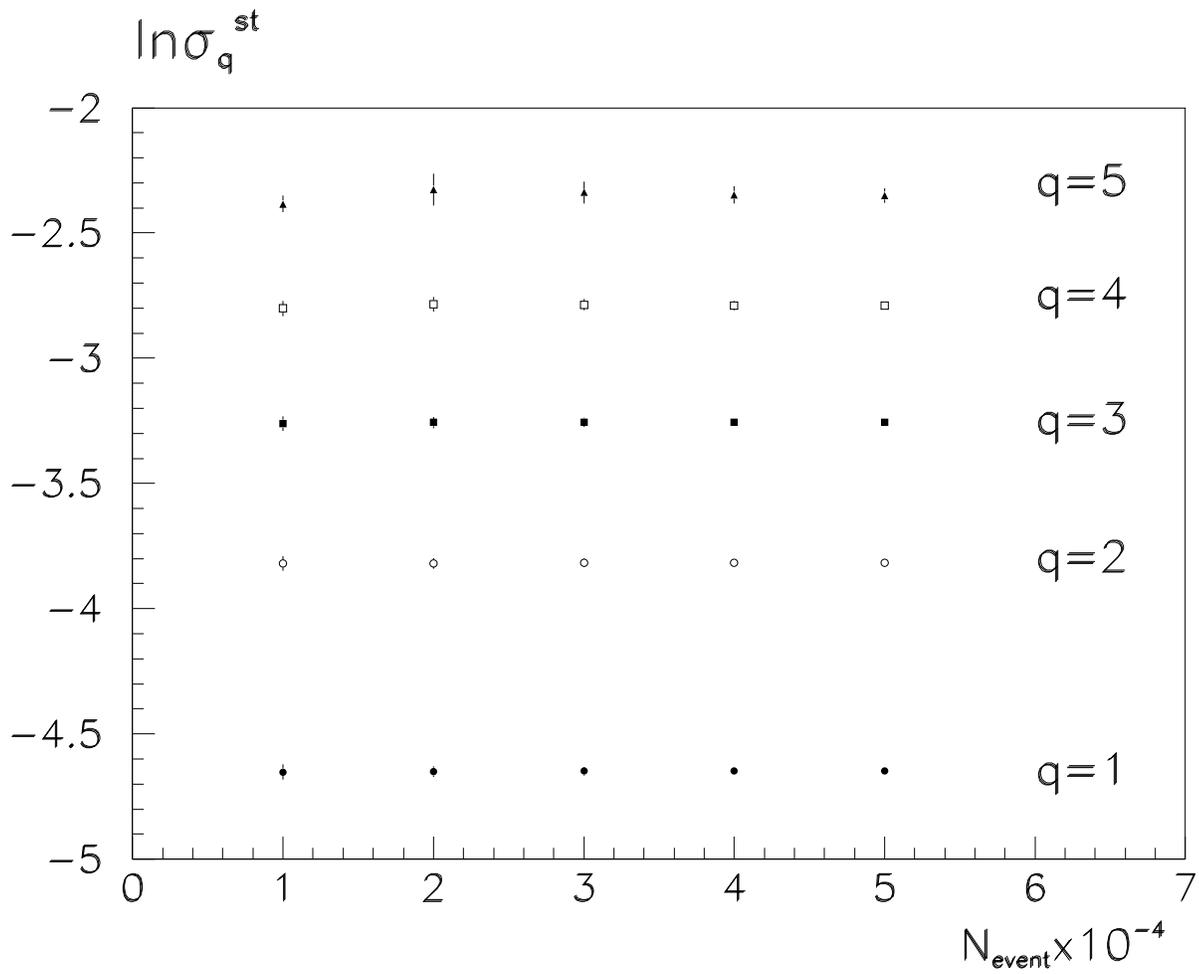
\item[\bf Fig.~3] $\ln \sigma_{q}^{st}$ ${\sl vs.}$ $N_{event}$ for different order of moments 
$q$ after shifting multiplicity $n$ to $11-35$, where the full circles, open circles, full 
squares, open squares and full triangles represent $q=1, 2, 3, 4, 5$ respectively.

\end{itemize}

\end{document}